\pdfoutput=1

\documentclass[11pt]{article}

\usepackage[preprint]{acl}

\usepackage{times}
\usepackage{latexsym}
\usepackage{amsmath, amssymb}
\usepackage{algorithm, algorithmicx, algpseudocode}

\usepackage[T1]{fontenc}
\usepackage[utf8]{inputenc}

\usepackage{microtype}
\usepackage{adjustbox}
\usepackage{multirow}
\usepackage{booktabs}
\usepackage{inconsolata}

\usepackage{stfloats}

\usepackage{graphicx}
\usepackage{subcaption}
\usepackage{authblk}
\usepackage{enumitem}
\newcommand{\AlgName}{FairKV}

\title{FairKV: Balancing Per-Head KV Cache for Fast Multi-GPU Inference}

\author{  
  
  \textbf{Bingzhe Zhao\textsuperscript{1*}},
  \textbf{Ke Cheng\textsuperscript{2*}},
  \textbf{Aomufei Yuan\textsuperscript{1}},
  \textbf{Yuxuan Tian\textsuperscript{1}},
  \textbf{Ruiguang Zhong\textsuperscript{3}},
  \vspace{-10pt}
  \\ 
  \textbf{Chengchen Hu\textsuperscript{3}},
  \textbf{Tong Yang\textsuperscript{1}},
  \textbf{Lian Yu \textsuperscript{1}}
  \\
  \textsuperscript{1}Peking University,
  \textsuperscript{2}University of Science and Technology of China,
  \textsuperscript{3}NIO Inc.,\\

}

\begin{document}
\maketitle
\renewcommand{\thefootnote}{\fnsymbol{footnote}} % 必须放在 \maketitle 之后
\footnotetext[1]{These authors contributed equally to this work.} % 正确匹配符号系统的索引
%\renewcommand{\thefootnote}{\fnsymbol{footnote}}
%\footnotetext[1]{These authors contributed equally to this work.}

%\input{0.abstract}
\begin{abstract}   
%KV cache techniques in Transformer models trade increased memory usage for reduced redundant computation. However, this approach leads to huge GPU memory usage, making KV cache compression an important and popular research topic. 
%
KV cache techniques in Transformer models aim to reduce redundant computations at the expense of substantially increased memory usage, making KV cache compression an important and popular research topic.
Recently, many popular  KV cache compression methods implement imbalanced, per-head allocation algorithms that dynamically adjust the KV cache budget for each attention head, achieving excellent performance in single-GPU scenarios. However, we observe that such imbalanced compression leads to significant load imbalance when deploying multi-GPU inference, as some GPUs become overburdened while others remain underutilized.
In this paper, we propose \AlgName{}, a method designed to ensure fair memory usage among attention heads in systems employing imbalanced KV cache compression. 
The core technique of \AlgName{} is \emph{Fair-Copying}, which replicates a small subset of memory-intensive attention heads across GPUs using data parallelism to mitigate load imbalance. 
Theoretical analysis provides insights, while experiments on popular models, including LLaMA 70b and Mistral 24b model, demonstrate that \AlgName{} increases throughput by 1.66x compared to standard tensor parallelism inference. Our code will be released as open source upon acceptance.
\end{abstract}
\section{Introduction}

\subsection{Background and Motivation}

Large-scale Transformer-based models are at the core of modern artificial intelligence. To support fast inference, these models rely on a key-value (KV) cache\cite{Vaswani2017AttentionIA,Dai2019TransformerXLAL,Rae2019CompressiveTF} that stores key and value embeddings from previously generated tokens, trading memory usage for reduced redundant computation. This cache prevents redundant computations and is crucial for efficient sequence generation. 
%
%However, the huge memory usage of KV cache becomes a bottleneck of inference system, and extensive works study how to compress KV cache\cite{引用7篇左右论文}.
%
However, the huge memory usage of the KV cache becomes a bottleneck for inference systems, and a great number works have investigated KV cache compression\cite{ge2024modeltellsdiscardadaptive,Liu2023CacheGenKC,xiao2024efficientstreaminglanguagemodels,li2024snapkvllmknowslooking,cai2024pyramidkvdynamickvcache,feng2024adakv,fu2024headsmatterheadlevelkv,Adnan2024KeyformerKC}.

%Traditional KV cache compression methods, such as StreamingLLM\cite{xiao2024efficientstreaminglanguagemodels}, SnapKV\cite{li2024snapkvllmknowslooking}, PyramidKV\cite{cai2024pyramidkvdynamickvcache}, allocate an equal/fair KV cache budget by default to all attention heads. 
%
Traditional KV cache compression methods (StreamingLLM\cite{xiao2024efficientstreaminglanguagemodels}, SnapKV\cite{li2024snapkvllmknowslooking} and PyramidKV\cite{cai2024pyramidkvdynamickvcache}) naturally allocate an equal (fair) KV cache budget to all attention heads.
This fair allocation simplifies the design and ensures uniform memory usage across heads. 
%
%Recently, SOTA works of KV cache such as AdaSnapKV\cite{feng2024adakv} and HeadKV\cite{fu2024headsmatterheadlevelkv} introduced imbalanced (unfair) per-head KV cache compression algorithms that dynamically adjust the KV cache budget for each attention head. 
%
Recently, many popular works of KV cache compression methods (AdaSnapKV\cite{feng2024adakv} and HeadKV\cite{fu2024headsmatterheadlevelkv}) introduce \textit{imbalanced per-head KV cache compression algorithms} that dynamically adjust the KV cache budget for each attention head.
%These methods compress the KV cache more effectively and achieve the SOTA performance for scenarios only considering one GPU.
%
These methods compress the KV cache more effectively and achieve impressive performance in single-GPU scenarios.

%\textbf{Problem:} Our investigation reveals that the imbalanced per-head KV cache compression techniques, though beneficial for reducing overall memory usage, lead to \textbf{unfair load distribution} when deploying multi-GPU inference. 
%
% \noindent\textbf{Unfair head load Problem:} Our investigation reveals that although imbalanced per-head KV cache compression techniques reduce overall memory usage, they result in an \textit{Unfair per-Head KV cache Load distribution} during multi-GPU inference, and we named it \textbf{unfair head load} for convenience.
% %
% In systems employing tensor parallelism (one of the most common parallel strategy), this non-uniform KV cache usage forces some GPUs to handle a disproportionate share of memory-intensive attention heads.
% This imbalance (unfairness) increases GPU idle time, elevates inference latency, and ultimately degrades system throughput. 
% %
% Despite considerable research on load balancing, no prior work has identified or addressed this specific head-level imbalance caused by imbalanced KV cache compression. 
% %
% Tackling this problem is critical for achieving efficient, low-latency inference in large-scale Transformer models.

\noindent\textbf{Unfair head load Problem:} 
Our investigation reveals that although imbalanced per-head KV cache compression techniques reduce overall memory usage, they result in an \textit{uneven per-head KV cache load distribution} during multi-GPU inference, which we refer to as the \textbf{unfair head load} problem. In systems using tensor parallelism, one of the most common parallel strategies, this non-uniform KV cache usage forces some GPUs to handle a disproportionate share of memory-intensive attention heads. This imbalance increases GPU idle time, elevates inference latency, and ultimately degrades system throughput.
Despite extensive research on load balancing in parallel inference systems, no prior work has identified or addressed this specific imbalance caused by imbalanced KV cache compression. Addressing this imbalance is crucial for achieving efficient, low-latency inference in large-scale Transformer models, particularly in multi-GPU settings.

\subsection{Our solution}

%\subsubsection{Overview}
To address the unfair head load problem, we propose \AlgName{}. Our approach targets the fair memory usage among attention heads in imbalanced KV cache compression systems. \AlgName{} mainly employs two techniques to address this issue: best-effort assignment and fair-copying.
 \textit{Best-effort Assignment} is responsible for distributing attention heads across GPUs to achieve a relatively balanced workload.
%An \textbf{advanced version}, termed \emph{Fair-Copying}, which replicates a small subset of memory-intensive attention heads across GPUs using data parallelism, further mitigating load imbalance. 
 \textit{Fair-Copying} involves replicating certain attention heads to participate in the assignment, thereby reducing the workload of the replicated heads.
%
%An \textbf{advanced version}, namely \emph{Fair-Copying}, which selectively replicates a small subset of memory-intensive attention heads across GPUs using data parallelism, further mitigating load imbalance.
%

%\subsubsection{Basic Version: Best-effort Assignment}
%\noindent\textbf{Basic Version: Best-effort Assignment}
\noindent\textbf{Technique I: Best-effort Assignment.}
Best-effort Assignment works as follows.
We first analyze the KV cache consumption of each attention head under imbalanced compression. Based on this analysis, our allocation algorithm assigns attention heads to GPUs such that the aggregate memory and computational load is balanced across GPUs in a best-effort manner. This method requires no additional overhead and can be integrated into existing multi-GPU inference systems easily, offering a practical improvement in load balancing without modifying the underlying KV cache compression scheme.
%
%Best-effort Assignment works as follows. We analyze the KV cache consumption of each attention head under imbalanced compression and assign heads to GPUs using a simple heuristic. This best-effort assignment minimizes load variance without adding extra overhead, providing an immediate but limited improvement in load balancing.

%\noindent\textbf{Advanced Version: Fair-Copying}
\noindent\textbf{Technique II: Fair-Copying.}
While best-effort assignment offers a straightforward solution, some attention heads remain significantly more memory-intensive, leading to persistent bottlenecks. 
To further improve load balance, we propose another technique called \emph{Fair-Copying}, which utilizes Data Parallel techniques to enhance the effectiveness of the assignment.
In this technique, we set a replication budget that allows the algorithm to attempt replicating attention heads. By utilizing data parallelism to reduce the load on the replicated heads, these replicated heads participate in the assignment alongside the original ones, thereby expanding the search space and enabling finer-grained partitioning. This replication method minimizes additional overhead while substantially reducing GPU idle time and inference latency.

\noindent\textbf{Key Contributions.}
This paper makes the following key contributions: 
\begin{enumerate} [label=\arabic*)]
\vspace{-0.1in}
\item To the best of our knowledge, we are the first to reveal the unfair head load problem: imbalanced per-head KV cache compression causes significant load imbalance in multi-GPU inference environments.

%\item We propose \AlgName{}, which includes two mechanisms: best-effort assignment and fair-copying, to balance the workload across GPUs for imbalanced per-head KV cache compression systems.
\vspace{-0.1in}
\item To address the unfair head load problem,
we propose \AlgName{}, which includes two mechanisms—best-effort assignment and fair-copying.

%\vspace{-0.1in}
%\item We develop a mathematical model and an optimization algorithm to guide the assignment and replication of attention heads. 

\vspace{-0.1in}
\item Theoretical analysis provides insights, while experiments on popular models, including LLaMA 70b and Mistral 24b model, demonstrate that \AlgName{} increases throughput by 1.66x compared to standard tensor parallelism inference. Our code will be released as open source upon acceptance.
\end{enumerate}
\section{Related Work}

Inference efficiency for large language models is critically limited by both memory bandwidth and computational power. KV cache compression \cite{ge2024modeltellsdiscardadaptive, zhang2024h2o, yang2024pyramidinfer}, a widely adopted optimization technique, reduces memory for storing previous key-value states in attention layers. It can generally be classified into two categories: \textit{Balanced (Fair) Per-Head Compression} and \textit{Imbalanced (Unfair) Per-Head Compression}.

\textit{Balanced (Fair) Per-Head Compression} methods applies the same strategy to all attention heads. For instance, StreamingLLM \cite{xiao2024efficientstreaminglanguagemodels} retains the initial $k$ sink tokens along with the recent window. H2O \cite{zhang2024h2o} further prioritizes important cache entries based on accumulated attention scores, while SnapKV \cite{li2024snapkvllmknowslooking} selects entries using attention scores from the observation window. Recently, Pyramid \cite{yang2024pyramidinfer, cai2024pyramidkvdynamickvcache} recognize distinct attention distribution patterns across different layers. However, these methods disregard the varying importance of different heads in actual computations. 

In contrast, \textit{Imbalanced (Unfair) Per-Head Compression} algorithms, like Ada-SnapKV \cite{feng2024adakv} and HeadKV\cite{fu2024headsmatterheadlevelkv}, dynamically adjust the KV cache budget per attention head based on the current layer's computational and memory requirements. Ada-SnapKV determines the budget for each head during inference, with a fully dynamic allocation. HeadKV, however, pre-allocates a fixed base budget for each head according to its importance and then adds a dynamic budget. This tailored cache allocation offers more flexibility and optimization potential. Table~\ref{tab:KV_Methods} in the appendix show that Ada-SnapKV outperforms other methods on multiple tasks in the LongBench v1 \cite{bai2024longbenchbilingualmultitaskbenchmark}, demonstrating the effectiveness of \textit{Imbalanced (Unfair) Per-Head Compression} approach.

However, a significant challenge arises when applying \textit{Imbalanced (Unfair) Per-Head Compression} in tensor parallelism, which has become the preferred method for inference in large language models \cite{lu2017flexflow, shazeer2018mesh, shoeybi2019megatron, rajbhandari2020zero}. The non-uniform distribution of KV cache across heads causes computational load imbalance, degrading overall inference efficiency. This highlights the urgent need for balancing per-head KV cache.

% \begin{figure*}[htbp]
%     \centering
%     \includegraphics[width=\textwidth]{latex/KV_methods_with_TP.png}
%     \caption{Common KV Cache compression methods with Tensor Parallelism}
%     \label{fig:KV_methods_with_TP}
% \end{figure*}
\begin{table*}[ht]
    \centering
    \caption{Cosine Similarity of Retained KV Cache in Per-Head KV Compression across LongBench Sub-datasets}
    \fontsize{15}{14}\selectfont
    \renewcommand{\arraystretch}{1.2} % 调整行间距
    \setlength{\tabcolsep}{4pt} % 调整列间距
    \begin{adjustbox}{max width=\textwidth}
    \begin{tabular}{ccccccccccccccccccc}
        \toprule
        \multirow{2}{*}{KV Budget} & 
        \multicolumn{3}{c}{Single-Doc QA} & 
        \multicolumn{3}{c}{Multi-Doc QA} & 
        \multicolumn{3}{c}{Summarization} & 
        \multicolumn{3}{c}{Few-shot Learning} & 
        \multicolumn{2}{c}{Coding} & 
        \multirow{2}{*}{Avg} &
        \multirow{2}{*}{Max} &
        \multirow{2}{*}{Min} &
        \multirow{2}{*}{Std}\\
        \cmidrule(lr){2-4} \cmidrule(lr){5-7} \cmidrule(lr){8-10} \cmidrule(lr){11-13} \cmidrule(lr){14-15}
        & NtrQA & Qasper & MF-en & HpQA & 2WMQA & Musiq & GovRp & QMSum & MNews & TREC & TriQA & SAMSum & LCC & RB-P \\
        \midrule
        \multicolumn{19}{c}{Llama-3.3-70B-Instruct } \\
        128 & 0.974 & 0.977 & 0.977 & 0.979 & 0.980 & 0.979 & 0.971 & 0.979 & 0.974 & 0.974 & 0.978 & 0.969 & 0.975 & 0.977 & 0.976 &
        0.980 & 0.969 & 0.003\\
        256 & 0.961 & 0.965 & 0.967 & 0.970 & 0.971 & 0.969 & 0.958 & 0.966 & 0.962 & 0.962 & 0.968 & 0.949 & 0.964 & 0.968 & 0.964 &
        0.971 & 0.949 & 0.006\\
        512 & 0.955 & 0.962 & 0.962 & 0.965 & 0.968 & 0.964 & 0.954 & 0.960 & 0.955 & 0.957 & 0.964 & 0.944 & 0.959 & 0.965 & 0.959 &
        0.968 & 0.944 & 0.006\\
        1024 & 0.950 & 0.961 & 0.959 & 0.961 & 0.966 & 0.959 & 0.953 & 0.956 & 0.943 & 0.956 & 0.962 & 0.947 & 0.949 & 0.962 & 0.956 &
        0.966 & 0.943 & 0.006\\
        \midrule
        \multicolumn{19}{c}{Meta-Llama-3-8B } \\
        128 & 0.904 & 0.935 & 0.932 & 0.919 & 0.933 & 0.914 & 0.929 & 0.919 & 0.938 & 0.919 & 0.928 & 0.934 & 0.933 & 0.932 & 0.926 &
        0.938 & 0.904 & 0.009\\
        256 & 0.873 & 0.912 & 0.913 & 0.897 & 0.917 & 0.887 & 0.912 & 0.891 & 0.923 & 0.896 & 0.909 & 0.911 & 0.915 & 0.909 & 0.905 &
        0.923 & 0.873 & 0.013\\
        512 & 0.886 & 0.916 & 0.917 & 0.909 & 0.924 & 0.902 & 0.917 & 0.904 & 0.933 & 0.909 & 0.917 & 0.908 & 0.925 & 0.915 & 0.913 &
        0.933 & 0.886 & 0.011\\
        1024 & 0.912 & 0.932 & 0.934 & 0.930 & 0.937 & 0.930 & 0.931 & 0.927 & 0.949 & 0.929 & 0.935 & 0.922 & 0.944 & 0.932 & 0.932 &
        0.949 & 0.912 & 0.009\\
        \midrule
        \multicolumn{19}{c}{
Mistral-Small-24B-Instruct-2501 } \\
        128 & 0.970 & 0.973 & 0.974 & 0.974 & 0.975 & 0.973 & 0.971 & 0.975 & 0.970 & 0.959 & 0.975 & 0.971 & 0.971 & 0.973 & 0.972 &
        0.975 & 0.959 & 0.004\\
        256 & 0.960 & 0.963 & 0.965 & 0.965 & 0.967 & 0.965 & 0.959 & 0.965 & 0.960 & 0.944 & 0.967 & 0.951 & 0.960 & 0.965 & 0.961 &
        0.967 & 0.944 & 0.006\\
        512 & 0.954 & 0.958 & 0.960 & 0.958 & 0.964 & 0.958 & 0.951 & 0.959 & 0.959 & 0.945 & 0.963 & 0.936 & 0.955 & 0.962 & 0.956 &
        0.964 & 0.936 & 0.007\\
        1024 & 0.954 & 0.961 & 0.961 & 0.959 & 0.965 & 0.957 & 0.951 & 0.959 & 0.963 & 0.954 & 0.962 & 0.937 & 0.959 & 0.962 & 0.957 &
        0.965 & 0.937 & 0.007\\
        \bottomrule
    \end{tabular}
    \end{adjustbox}
    \label{tab:longbench}
\end{table*}

\begin{table}[ht]
\centering
\caption{GPU Busy Rate of Different Models. We applied Ada-SnapKV to three models, setting the KV cache budget to 128, 256, 512, and 1024, and measured its GPU busy rate under tensor parallel sizes of 2, 4, and 8.}
\fontsize{24}{22}\selectfont
\renewcommand{\arraystretch}{1.2} % 调整行间距
\setlength{\tabcolsep}{4pt} % 调整列间距
\label{tab:SHA GPUutilization}
\resizebox{\columnwidth}{!}{%
\begin{tabular}{c c c c c}
\toprule
Model & KV Cache Budget & TP = 2 & TP = 4 & TP = 8 \\
\midrule
\multirow{4}{*}{LLaMA-3.3-70B-Instruct} 
    & 128  & 92.5 & 81.6 & 64.7 \\
    & 256  & 87.5 & 74.1 & 57.2 \\
    & 512  & 86.4 & 70.5 & 55.7 \\
    & 1024 & 87.2 & 71.2 & 58.2 \\
\midrule
\multirow{4}{*}{Meta-LLaMA-3-8B} 
    & 128  & 92.1 & 84.4 & 70.8 \\
    & 256  & 91.8 & 82.0 & 68.5 \\
    & 512  & 90.6 & 81.6 & 68.8 \\
    & 1024 & 90.9 & 82.1 & 69.8 \\
\midrule
\multirow{4}{*}{Mistral-Small-24B-Instruct} 
    & 128  & 93.2 & 86.3 & 75.2 \\
    & 256  & 91.7 & 82.9 & 71.9 \\
    & 512  & 91.2 & 82.0 & 70.8 \\
    & 1024 & 91.2 & 82.1 & 70.8 \\
\bottomrule
\end{tabular}%
}
\end{table}

\begin{figure*}[ht]
    \centering
    \begin{subfigure}[b]{0.40\linewidth}    % Slightly increased width
        \includegraphics[width=\linewidth]{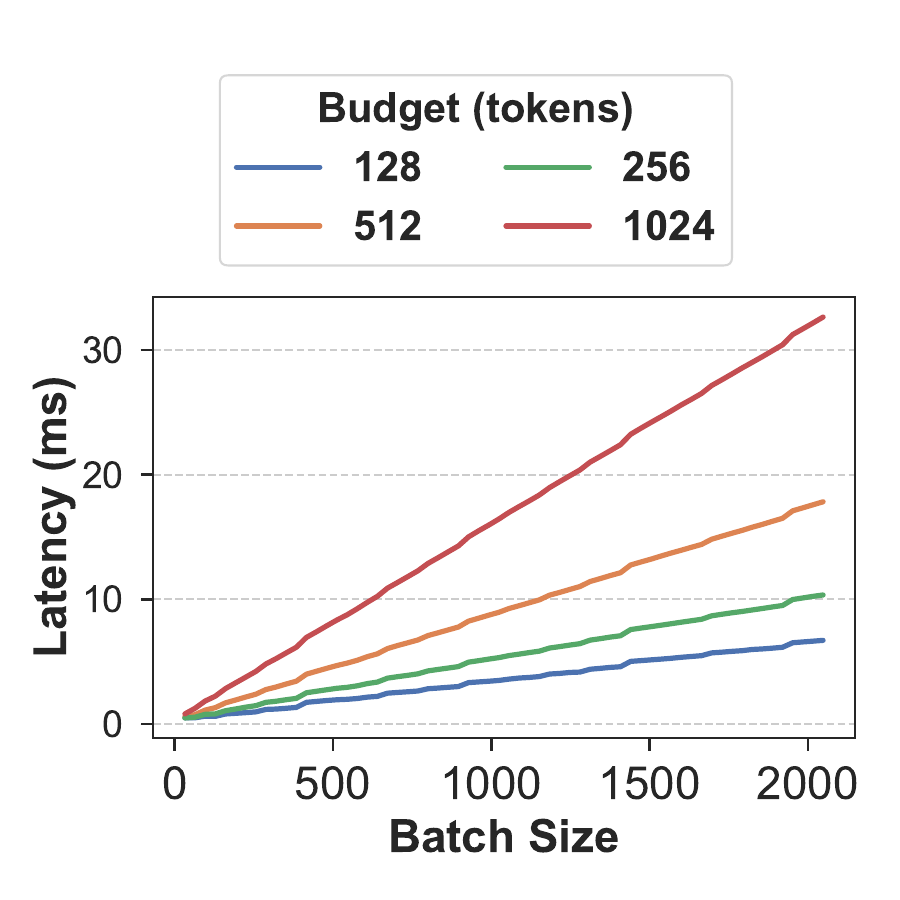}
        \caption{Latency vs. Batch Size under Different Budgets}
        \label{fig:image-a}
    \end{subfigure}
    % \hfill
    \hspace{0.1cm}
    \begin{subfigure}[b]{0.40\linewidth}    % Slightly increased width
        \includegraphics[width=\linewidth]{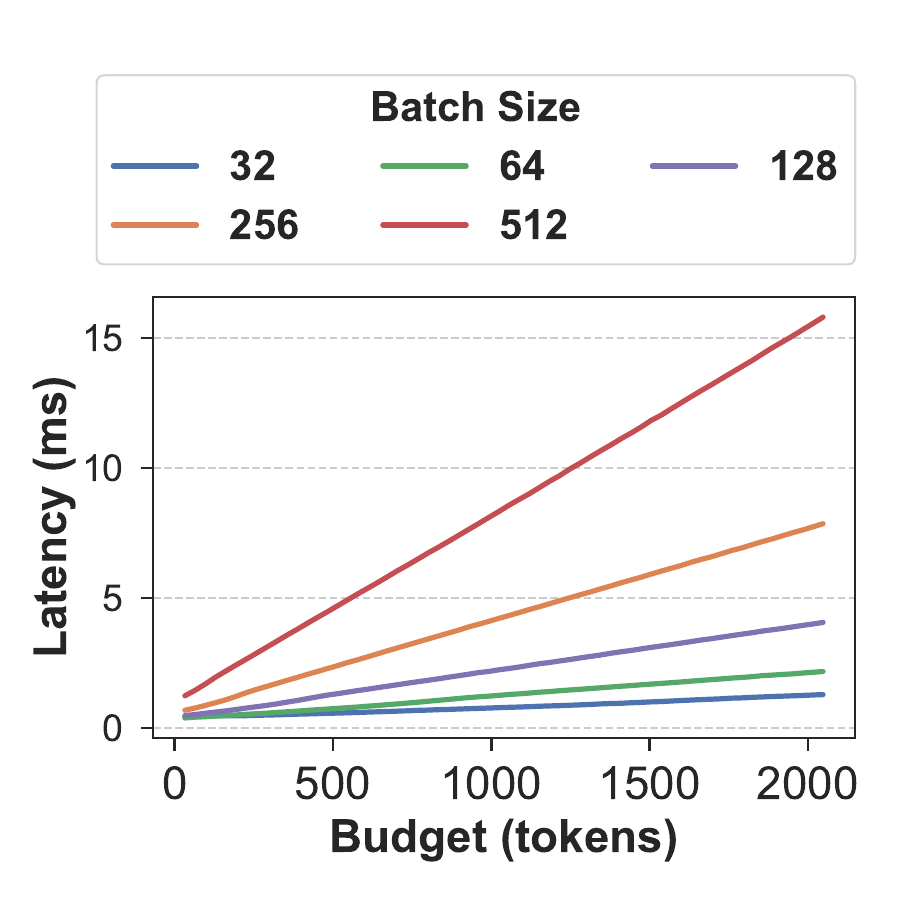}
        \caption{Latency vs. Budget under Different Batch Sizes}
        \label{fig:image-b}
    \end{subfigure}
    \caption{Impact of Batchsize and KV Cache Budget on Inference Latency}
    \label{fig:bs_bud_lat}
\end{figure*}
\section{Preliminary}

\subsection{Observation of KV Cache Selection}

We examined per-head KV cache eviction patterns across multiple datasets using different models (LLaMA-3.3-70B-Instruct, Meta-LLaMA-3-8B, and Mistral-Small-24B-Instruct) and several subsets from LongBench v1 for our experiments and set the KV budget for the attention heads to 128, 256, 512, and 1024 as our basic experimental setup.

The per-head KV cache compression algorithm results in varying KV cache budgets across different attention heads, leading to imbalanced workloads when combined with tensor parallelism.   We introduce GPU busy rate, defined as the ratio of time the GPU spends actively executing inference computations to the total latency of the entire inference process, to quantify the resource underutilization caused by workload imbalance during parallel execution. The results, summarized in Table~\ref{tab:SHA GPUutilization}, show a clear trend: 

Decreasing GPU Busy Rate with Larger TP Sizes:  For instance, in LLaMA-3.3-70B-Instruct with KV cache budget 128, GPU busy rate drops from 92.5\% (TP=2) to 81.6\% (TP=4) and further to 64.7\% (TP=8). A similar trend is observed in Meta-LLaMA-3-8B, where GPU utilization declines from 92.1\% (TP=2) to 84.4\% (TP=4) and 70.8\% (TP=8) under the same KV cache budget.

Impact of KV Cache Budget: Lower KV cache budgets generally yield better GPU busy rate. For example, in Mistral-Small-24B-Instruct with TP=4, GPU busy rate is 86.3\% at KV=128 but decreases to 82.9\% at KV=256 and further to 82.1\% at KV=1024. Similarly, in LLaMA-3.3-70B-Instruct with TP=4, GPU busy rate drops from 81.6\% (KV=128) to 71.2\% (KV=1024). This suggests that larger KV cache budgets lead to greater workload imbalance among attention heads, which negatively impacts GPU efficiency.  

Additionally, as shown in Table \ref{tab:longbench}, we use cosine similarity to quantify the difference in KV cache allocation between a subset and the remaining datasets. Finally, we averaged the metrics across all subsets to obtain an overall indicator. Our analysis indicates that eviction patterns remain largely consistent across datasets, confirming that the statistics for retained KV cache can guide optimization. Furthermore, the allocation pattern of the KV cache budget is influenced by the specific model used. Given this dataset-invariant nature, optimization strategies can be designed based on these profiles without requiring per-dataset recalibration.

\subsection{Empirical Performance Analysis}

To optimize the distribution of GPU workload, we built an empirical model mapping batch size, retained KV cache count, and inference latency. Measurements were taken across various configurations on a multi-GPU setup.

We conducted experiments on the LongBench v1 dataset by systematically varying both batch size and KV cache budget parameters. Specifically, we controlled these variables independently to evaluate their individual and combined effects. We simulated the inference scenario of LLaMA 70b on a single layer and obtained the latency during decoding one token. The experimental results, as illustrated in Figure \ref{fig:bs_bud_lat}, demonstrate the performance characteristics under different configurations.

For the batch size latency model (Figure a), we observe that latency ($L$) increases approximately linearly with batch size ($B$) across different budget configurations. This is because during the decoding phase of large model inference, due to the autoregressive characteristic of Transformer-based architectures, the inference latency is primarily determined by GPU bandwidth rather than computational power (memory-bound). Consequently, as the batch size increases, the data transfer volume grows, making data transmission the dominant bottleneck, and latency increases nearly linearly. The relationship can be expressed as $L \approx \alpha B + \beta$, where $\alpha$ represents the slope and $\beta$ the initial offset. The graph shows four distinct budget levels (128, 256, 512, and 1024), with higher budget values corresponding to steeper slopes in the linear relationship.

Similarly, the KV cache latency model (Figure b) demonstrates a comparable linear pattern, where latency ($L$) increases proportionally with the KV cache budget ($C$). This occurs since the expanded KV Cache budget also amplifies data transfer demands, thereby maintaining the memory-bound bottleneck that dominates latency behavior. This can be represented as $L \approx \gamma C + \delta$, where $\gamma$ and $\delta$ are the slope and offset parameters respectively. The data presents five different batch sizes (32, 64, 128, 256, and 512), with larger batch sizes showing more pronounced slopes in their linear relationships.

In both cases, the approximately linear nature of these relationships is particularly noteworthy, as it suggests predictable scaling behavior in the system's performance characteristics. Therefore, we conclude that during the decoding phase of model inference, neither increasing batch size nor expanding the KV Cache budget yields substantial efficiency improvements. This persistence of linear latency scaling fundamentally stems from the memory-bound nature of autoregressive generation in Transformer architectures, where GPU bandwidth constraints dominate computational throughput.
\section{Design of \AlgName{}}
\subsection{Overview}

\begin{figure*}[htbp]
  \centering
  \includegraphics[width=\textwidth]{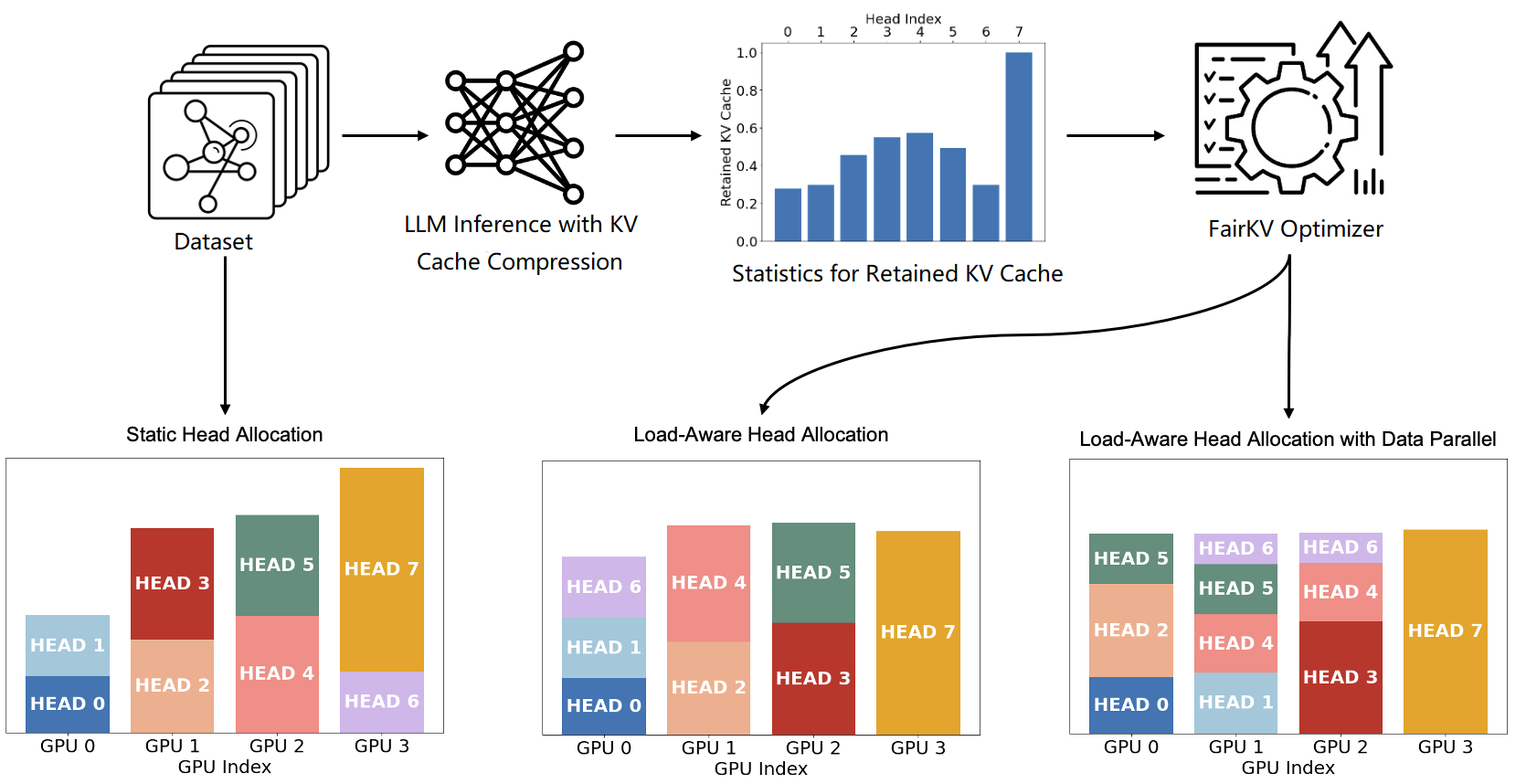} 
  \caption{Illustration of different head allocation strategies for multi-GPU inference in large-scale transformer models. The figure shows the following strategies: (1) Static Head Allocation (SHA), where attention heads are evenly distributed across GPUs without considering computational load; (2) Load-Aware Head Allocation (\AlgName{}-NoDP), where attention heads are allocated based on their computational load, ensuring a balanced GPU busy rate; (3) Load-Aware Head Allocation with DataParallel (\AlgName{}-DP), where heads are replicated across GPUs for improved load distribution and efficiency; }
  \label{fig:aha_overview}
\end{figure*}

During inference with tensor parallelism, per-head KV cache compression algorithms can lead to imbalanced GPU workloads, which in turn reduces inference efficiency. To address this issue, we designed \AlgName{}, a load-aware static approach that uses a search algorithm to reassemble and rearrange attention heads across layers, thereby balancing the load among GPUs. Moreover, \AlgName{} leverages fair-copying mechanism to expand the search space via replication of attention heads and DataParallel techniques, enabling a more fine-grained balancing of GPU loads and enhancing GPU busy rate during inference. Figure \ref{fig:aha_overview} visually illustrates the core concept of \AlgName{}.

The \AlgName{} algorithm requires predefining the model and the KV cache budget to be used. It then samples a dataset to analyze the proportion of KV cache budget allocated to attention heads across different layers for that model, summarizing the findings into a statistical profile. Based on this data, we first replicate attention heads to expand the search space. Next, we perform a constrained search to determine the optimal attention head arrangement. Finally, the model weights are loaded according to this arrangement for inference.

\subsection{Search Space}
Fair-copying leverages Data Parallel techniques to expand the search space by replicating some redundant heads, with the goal of achieving more effective search outcomes. Experiments were conducted with selective head replication, and the results demonstrated improved latency distribution across GPUs. Allowing for limited redundancy provides an effective way to enhance parallel efficiency without excessive computational overhead. To better illustrate FairKV's optimization objectives, we formulate a mathematical model, with detailed derivations provided in Appendix A.

\subsection{Optimizer Design}
We introduce a recursive backtracking algorithm that systematically explores possible head distributions and an optimal scheme selection algorithm to select the best attention head allocation scheme to promote workload balance.

We first use Algorithm~\ref{alg:backtrack} to enumerate all possible attention head replication schemes. Then, we apply a head-wise KV cache compression algorithm(like Ada-SnapKV) on a user-provided dataset to sample the KV cache allocation of each attention head per layer, which serves as the weight (denoted as W) for that layer. Finally, Algorithm~\ref{alg:selection} is employed for each layer to determine the optimal allocation scheme.

\begin{algorithm}
\caption{Backtracking-Based Partitioning Algorithm}\label{alg:backtrack}
\begin{algorithmic}[1]
\Require List of heads $L$, maximum copied heads $m$
\Ensure Result set $R$, all possible partitions of $L$
\State Initialize result set $R \gets \emptyset$
\Procedure{Backtrack}{$\mathit{index}$, $\mathit{p}$}
    \Statex \quad \Comment{$\mathit{p}$ is a possible division of L}
    \If{$\mathit{index} = |L|$}
        \State $R \gets R \cup \{\mathit{p}\}$ 
        \State \Return
    \EndIf
    
    \State Append $L[\mathit{index}]$ to $\mathit{p}$ 
    \Statex \quad \Comment{Case 1: Single copy}
    \State\Call{Backtrack}{$\mathit{index} + 1$, $\mathit{p}$}
    \State Remove last element from $\mathit{p}$
    
    \For{$n \gets 2$ \textbf{to} $|L| - \mathit{index} + |\mathit{p}|$ + 1} 
        \State Create $n$ copies of $L[\mathit{index}]$ as $L_n$
        \State Append values of $L_n$ to $\mathit{p}$  
        \Statex \quad \Comment{Case 2: Multiple copies}
        \State\Call{Backtrack}{$\mathit{index} + 1$, $\mathit{p}$}
        \State Remove values of  $L_n$  from $\mathit{p}$
    \EndFor
\EndProcedure
\Statex

\State \Call{Backtrack}{$0$, $\emptyset$}
\Statex \quad \Comment{Initial call with empty partition}
\State \Return $R$
\end{algorithmic}
\end{algorithm}

\begin{algorithm}
\caption{Optimal Scheme Selection}\label{alg:selection}
\label{alg:scheme_selection}
\begin{algorithmic}[1]
\Require $R$ from Algorithm 1, attention heads weight $W$ of one layer, tensor parallelism size $tp$
\Ensure best\_scheme, optimal allocation scheme of one layer
\State $\text{div\_value} \gets \infty$
\State $\text{best\_scheme} \gets [\ ]$
\For{each $p$ in $R$}
    \State $\text{copied\_size} \gets \Call{GetCopiedSize}{p}$ 
    \Statex \Comment{Get occurrence counts}
    \State $W_A \gets \{ W[i] / \!\text{copied\_size}[i] \ 
        | \ \forall i\}$ 
    \Statex \Comment{Adjust weights}
    \State $\mathcal{P} \gets \Call{SplitIntoTpGroups}{p, tp}$ 
    \Statex \Comment{Generate candidate schemes}
    \For{each $\text{scheme}$ in $\mathcal{P}$}
        \State $\Delta \gets \Call{MaxGroupWeight}{W_A} - 
            \Call{MinGroupWeight}{W_A}$
        \If{$\Delta < \text{div\_value}$}
            \State $\text{div\_value} \gets \Delta$
            \State $\text{best\_scheme} \gets \text{scheme}$
        \EndIf
    \EndFor
\EndFor
\State \Return $\text{best\_scheme}$
\end{algorithmic}
\end{algorithm}

Through these two algorithms, we can effectively improve GPU busy rates, thereby reducing idle waiting time across GPUs and enhancing inference efficiency.

\subsection{Key Innovations}

\noindent\textbf{Workload-aware Redistribution} A static yet optimized head allocation strategy balances GPU workloads efficiently.

\noindent\textbf{Hybrid Parallelism} Selective head replication combines tensor and data parallelism.

\noindent\textbf{Latency-driven Optimization} Backtracking minimizes inference latency, improving GPU busy rate.

\section{Evaluation}
In this section, we primarily conduct a detailed evaluation of the \AlgName{} method proposed in the previous sections, assessing whether \AlgName{} can effectively improve inference efficiency in hybrid parallelism approaches  during Per-Head KV cache Compression.

\begin{figure*}[t]
    \centering
    \begin{subfigure}{0.32\linewidth}
        \centering
        \includegraphics[width=\linewidth]{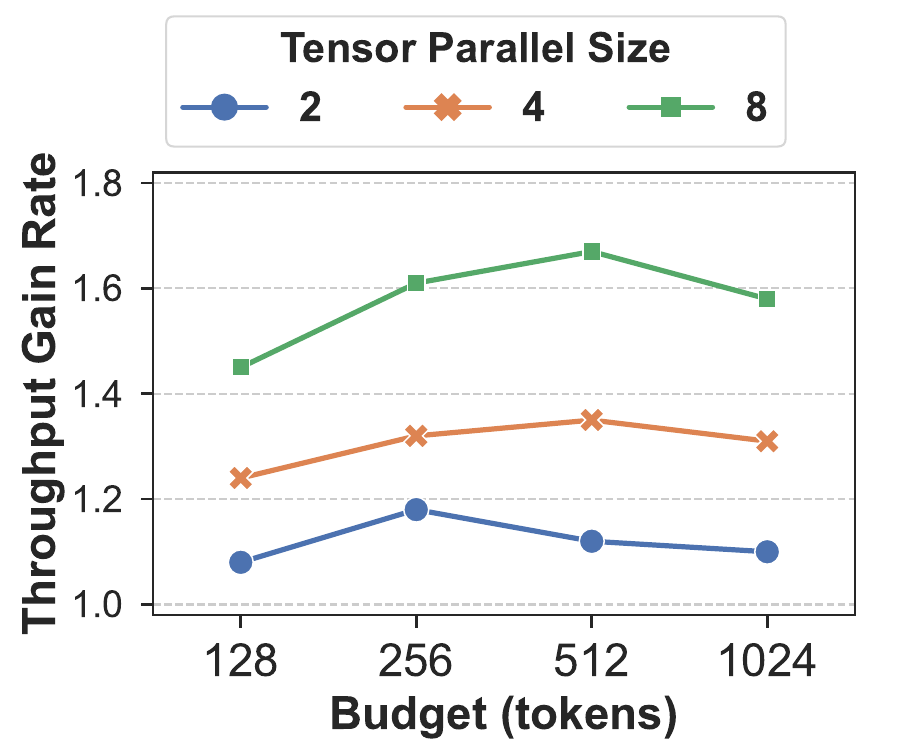}
        \caption{Rate on LLaMA-3.3-70B-Instruct}
        \label{fig:aha_tp16}
    \end{subfigure}
    \hfill
    \begin{subfigure}{0.32\linewidth}
        \centering
        \includegraphics[width=\linewidth]{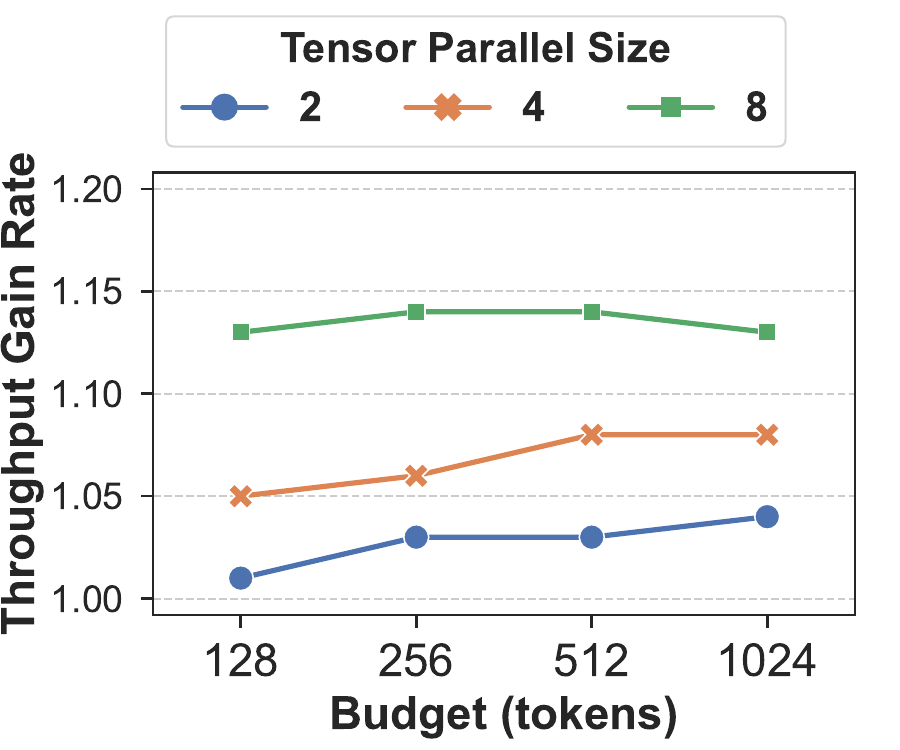}
        \caption{Rate on Meta-LLaMA-3-8B}
        \label{fig:aha_nodp}
    \end{subfigure}
    \hfill
    \begin{subfigure}{0.32\linewidth}
        \centering
        \includegraphics[width=\linewidth]{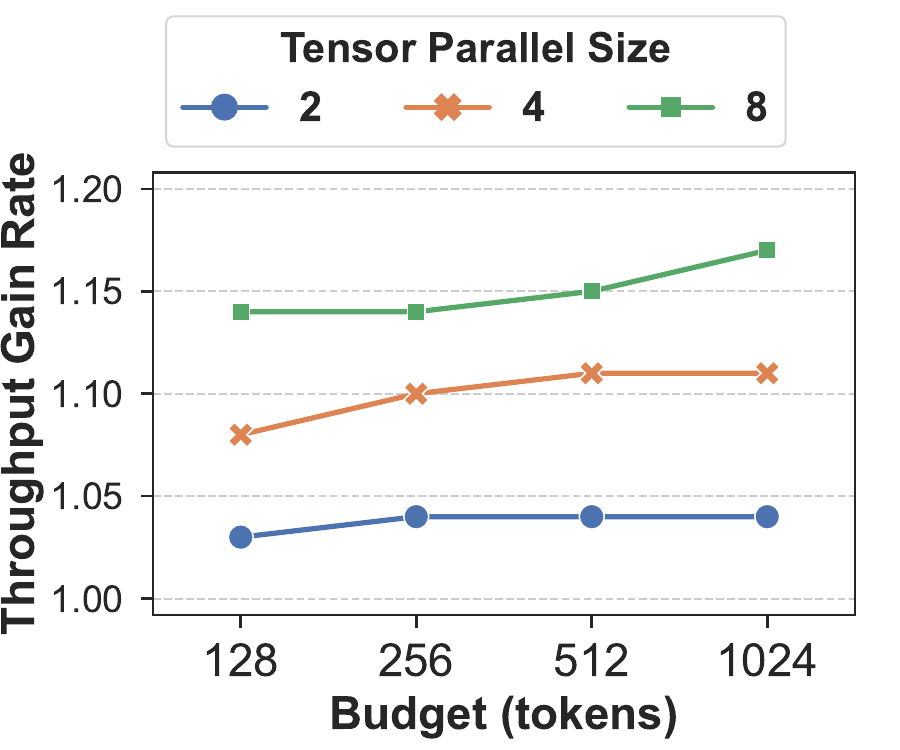}
        \caption{Rate on Mistral-Small-24B-Instruct}
        \label{fig:aha_dp}
    \end{subfigure}
    \caption{Throughput Gain Rates of \AlgName{} on different models, where the throughput of the baseline model is regarded as 1.0.}
    \label{fig:gainacrossmodel}
\end{figure*}

\subsection{Experimental Setup}
We used Python 3.10.16 as the programming language and implemented \AlgName{} on PyTorch. We leveraged AdaKV from KVPress as the per-head KV cache compression algorithm and tested our approach using Llama models \cite{llama3} of different sizes as well as the Mistral model \cite{jiang2023mistral7b}.

\textbf{Model and Hardware Configurations}  We used the LLaMA-3.3-70B-Instruct model, the Meta-LLaMA-3-8B model, and the Mistral-Small-24B-Instruct-2501 model as our experimental models. Our hardware environment consists of eightµ Nvidia A100-80G GPUs and 960GB of CPU memory.

\textbf{Dataset Configurations}
We used LongBench v1 \cite{bai2024longbenchbilingualmultitaskbenchmark} as the evaluation dataset. LongBench v1 is a bilingual, multi-task benchmark for long-text understanding, enabling a more rigorous evaluation of long-text comprehension.

\textbf{Baseline} To the best of our knowledge, this study is the first to highlight the issue of GPU workload imbalance caused by per-head KV cache compression algorithms in tensor parallelism. Therefore, we chose the situation of conducting tensor parallel inference with Per-Head KV Compression but without \AlgName{} as the baseline. By comparing the cases with and without \AlgName{}, we can determine whether \AlgName{} can improve GPU busy rate and reduce inference latency.

\textbf{Evaluation Metrics}
The purpose of the evaluation experiments is to verify the following two questions: First, whether \AlgName{} can effectively reduce inference latency, and second, whether \AlgName{} can effectively increase GPU busy rate. Therefore, our evaluation metrics are the following two aspects: inference time and GPU busy rate.

\subsection{Performance Evaluation}
Performance evaluation is a critical component in assessing the effectiveness of the \AlgName{} method in improving multi-GPU inference efficiency for large language models. This subsection outlines the key metrics and methodologies used to evaluate the performance of \AlgName{}, ensuring a comprehensive understanding of its impact on system efficiency.

\subsubsection{GPU busy rate Improvement}

\begin{figure}[ht]
  \centering
  \includegraphics[width=0.5\textwidth]{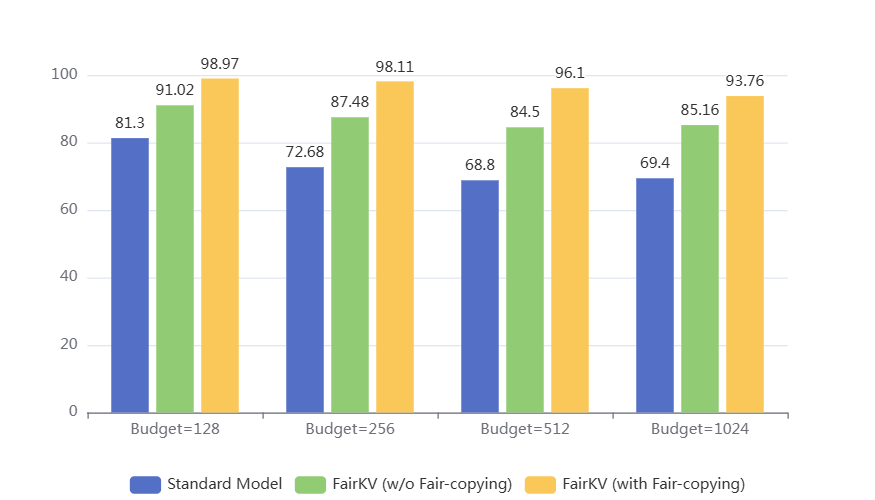} % 图片宽度设置为单列的适当比例
  \caption{Ablation Test Among Standard Model, \AlgName{} w/o Fair-copying and \AlgName{} with Fair-copying}
  \label{fig:ablation}
\end{figure}

\begin{figure}[ht]
  \centering
  \includegraphics[width=0.4\textwidth]{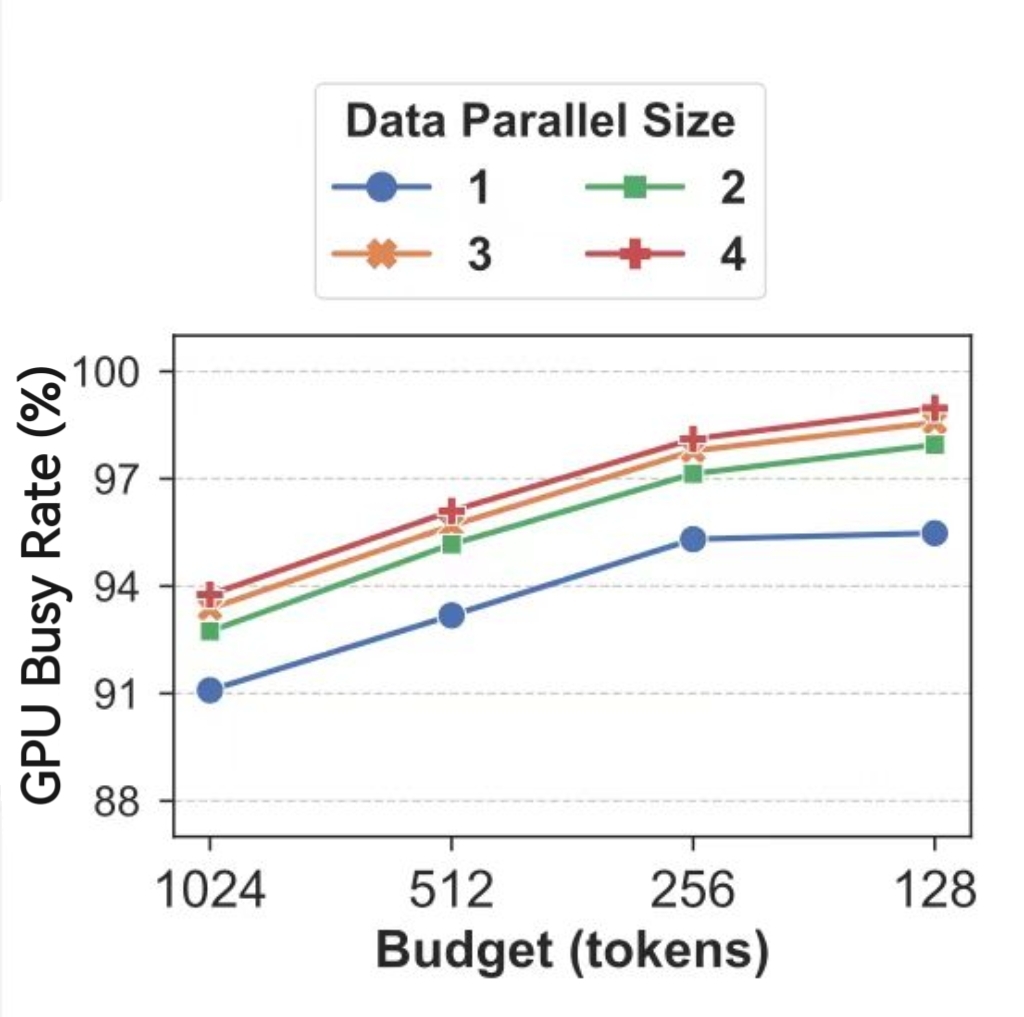} % 图片宽度设置为单列的适当比例
  \caption{GPU Busy Rate with different Data Parallel Size on LLaMA-3.3-70B.}
  \label{fig:GPUutilization}
\end{figure}

We used LLaMA-3.3-70B-Instruct to evaluate the impact of \AlgName{} on GPU busy rate. In Figure \ref{fig:ablation}, we conducted ablation tests among standard model, \AlgName{} without Fair-copying and \AlgName{} with Fair-copying. The result indicate that both \AlgName{} with or without Fair-copying significantly improves GPU busy rate compared to standard model, demonstrating that the \AlgName{} method can effectively balance GPU loads. Moreover, \AlgName{} with Fair-copying shows further improvement over \AlgName{} without Fair-copying. Then, to measure the impact of the parameter on the \AlgName{} with Fair-copying group, we set the size of parallel size, which is also the count of copied heads (CH), to 1, 2, 3, and 4. We measured the GPU busy rate for these groups under KV cache budgets of 128, 256, 512, and 1024, with the results shown in Figure \ref{fig:GPUutilization}.
The results suggesting that incorporating only a small number of copied heads via Data Parallel can enhance the performance of the \AlgName{} strategy greatly. We also observed a positive correlation between the performance of \AlgName{} and CH; as CH increases, the GPU busy rate curve becomes less steep, indicating that while increases in CH yield significant benefits when CH is small, the incremental gains diminish when CH is larger.

\subsubsection{Throughput Improvement}
We evaluated the performance of \AlgName{} across different models. Specifically, we selected a diverse set of models, including LLaMA-3.3-70B-Instruct, Meta-LLaMA-3-8B, and Mistral-Small-24B-Instruct, and set the tensor parallel size to either 4 or 8 with CH fixed at 4. Throughput gains relative to SHA were measured under KV cache budgets of 128, 256, 512, and 1024. As shown in Figure \ref{fig:gainacrossmodel}, \AlgName{} accelerates various models, and due to the inherent characteristics of each model, its throughput under SHA conditions differs. As a result, the acceleration effect of \AlgName{} may vary. Among them, \AlgName{} achieves the highest benefit of up to 1.66 on the Llama-3.3-70B-Instruct model.

We also evaluated the performance of \AlgName{} under different tensor parallel sizes. Firstly, we observed that \AlgName{} provides acceleration benefits under all tensor parallel size conditions. By comparing subfigures a, b, and c, we can see that as the tensor parallel size increases, the acceleration effect of \AlgName{} significantly improves under the same model and KV cache budget conditions. This suggests that \AlgName{} is more likely to achieve better acceleration performance when the tensor parallel size is larger.

As the KV cache budget increases, the acceleration effect of \AlgName{} shows a slight improvement for Meta-LLaMA-3-8B and Mistral-Small-24B-Instruct, while for LLaMA-3.3-70B-Instruct, the acceleration effect initially improves significantly and then slightly decreases. Overall, across these three models, the acceleration effect of \AlgName{} tends to improve as the KV cache budget increases.

In general, \AlgName{} demonstrates acceleration benefits across different models, tensor parallel sizes, and KV cache budgets. At the same time, the effectiveness of \AlgName{} is influenced by these factors, and the results may vary accordingly.

\section{Conclusion}

In this paper, we propose \textit{\AlgName{}}, a novel optimization technique designed to improve inference efficiency by dynamically adjusting the allocation of attention heads in Tensor Parallelism. \AlgName{} leverages the statistics of the retained Key-Value (KV) cache to partition attention heads based on their computational load, ensuring a balanced workload distribution across GPUs.

Our method has been extensively evaluated across various configurations, demonstrating consistent performance improvements. Specifically, \AlgName{} significantly reduces inference latency and improves GPU busy rate, regardless of the number of GPUs used, the model size (including different versions of LLaMa and Mistral), or the KV cache budget. This versatility highlights the robustness of our approach, making it effective across different settings and resource constraints. Experimental results show that \AlgName{} achieves up to a 66\% increase in inference throughput, with minimal impact on model accuracy. These findings confirm that \AlgName{} is a promising solution for optimizing large-scale model inference in real-world, real-time applications.

Future work will explore the adaptability of \AlgName{} to dynamic KV cache budgets and investigate its potential for scaling to models such as Mixture of Experts (MoE), where the computational load can vary significantly across different tasks.

\section{Limitations}

Although \textit{\AlgName{}} provides significant improvements in inference efficiency, there are a few limitations that need to be addressed. First, \AlgName{} is designed for a single machine with multiple GPUs and does not account for scenarios involving distributed systems across multiple machines. Second, the current method primarily focuses on the parallelization of the inference process without considering the separation of the prefill and decode stages, which is critical in certain real-time applications where the inference process is more complex and involves sequential decoding. 
Additionally, \AlgName{} relies on the accuracy of the KV cache statistics for optimal head allocation. This means that the effectiveness of KV cache compression depends on the statistical properties of the cache, and any deviations from these properties could reduce the method’s performance.

\clearpage

\bibliography{latex/acl_latex}

\clearpage

\appendix

\section{Mathematical Model for Hybrid
Parallelism in FairKV}
\subsection{Mathematical Model for Hybrid Parallelism in \AlgName{}}
This section presents a comprehensive mathematical formulation of the \AlgName{} system for optimizing multi-GPU inference in large language models with KV cache compression.

\subsubsection{System Parameters}
The system parameters are defined to model the components involved in the \AlgName{} strategy, for each transformer layer $l \in L$:
\begin{itemize}
    \item $H_l = \{h_1, ..., h_n\}$: attention heads
    \item $G = \{g_1, ..., g_m\}$: available GPUs
    \item $w_i$: workload for head $h_i$
    \item $r_{ij}$: replication factor of head $h_i$ on GPU $g_j$
\end{itemize}

\subsubsection{Decision Variables}
The decision variables are crucial for determining the optimal assignment of attention heads to GPUs:

\begin{equation}
    x_{ij} = \begin{cases} 
        r_{ij} & \text{if head } h_i \text{ is assigned to GPU } g_j \\
        0 & \text{otherwise}
    \end{cases}
\end{equation}

\subsubsection{Parallelism Constraints}

The hybrid parallelism approach in \AlgName{} combines Tensor Parallelism and Data Parallelism to address the challenges posed by head-wise KV Cache compression. Tensor Parallelism ensures that the computational workload is distributed across multiple GPUs by partitioning the attention heads, while Data Parallelism introduces redundancy by allowing selective replication of attention heads across GPUs. This hybrid approach aims to balance the computational load, ensuring that no single GPU becomes a bottleneck due to uneven KV Cache compression rates.

\textbf{Head Distribution}
Each head must have at least one GPU assignment to ensure that all computations are accounted for.
\begin{equation}
    \sum_{j \in G} \mathbb{1}(x_{ij} > 0) \geq 1 \quad \forall i \in H_l, \forall l \in L
\end{equation}

\textbf{Data Parallelism}
Total replication factor for each head:
\begin{equation}
    \sum_{j \in G} r_{ij} \leq R_{max} \quad \forall i \in H_l, \forall l \in L
\end{equation}
The total replication factor for each head across all GPUs must not exceed a predefined maximum, $R_{max}$. This prevents over-replication, which could lead to increased communication overhead and reduced efficiency.

\subsubsection{Optimization Objective}
The primary goal of the \AlgName{} strategy is to minimize the maximum processing time across all GPUs. This is formulated as:
\begin{equation}
    \min \max_{j \in G} \sum_{l \in L} \sum_{i \in H_l} \frac{x_{ij} w_i}{r_{ij}}
\end{equation}

\subsubsection{System Efficiency}
System efficiency is calculated to evaluate how well the GPUs are utilized. The formula for efficiency is:
\begin{equation}
    E = \frac{1}{|G|} \sum_{j \in G} \frac{\sum_{l \in L} \sum_{i \in H_l} \frac{x_{ij} w_i}{r_{ij}}}{\max_{k \in G} \sum_{l \in L} \sum_{i \in H_l} \frac{x_{ik} w_i}{r_{ik}}}
\end{equation}

\section{Value of Per-Head KV Cache Compression Methods}

Table~\ref{tab:KV_Methods} presents the performance of common KV cache compression methods on the LongBench v1 \cite{bai2024longbenchbilingualmultitaskbenchmark} dataset. We tested different KV cache compression methods on Llama-3.1-8B-Instruct with KV Cache budgets set to 128, 256, 512, 1024, and 2048. As described in Table~\ref{tab:KV_Methods}, across various KV cache budget settings, Ada-SnapKV (the optimized version of SnapKV based on Ada-KV) achieved higher average scores than other methods on multiple tasks in the Longbench v1 dataset.

\begin{table*}[htbp]
    \centering
    \renewcommand{\arraystretch}{1.1} 
    \setlength{\tabcolsep}{4pt} 
    \begin{adjustbox}{max width=\textwidth}
    \begin{tabular}{ccccccccccccccccccc}
        \toprule
        \multirow{2}{*}{Method} & 
        \multicolumn{3}{c}{Single-Doc QA} & 
        \multicolumn{3}{c}{Multi-Doc QA} & 
        \multicolumn{3}{c}{Summarization} & 
        \multicolumn{3}{c}{Few-shot Learning} & 
        \multicolumn{2}{c}{Synthetic} & 
        \multicolumn{2}{c}{Coding} & 
        \multirow{2}{*}{Ave. Score} \\
        \cmidrule(lr){2-4} \cmidrule(lr){5-7} \cmidrule(lr){8-10} \cmidrule(lr){11-13} \cmidrule(lr){14-15}  \cmidrule(lr){16-17}
        & NtrQA & Qasper & MF-en & HotpotQA & 2WikiMQA & Musique & GovReport & QMSum & MultiNews & TREC & TriviaQA & SAMSum & PCount & PRe & LCC & RB-P \\
        \midrule
        \multicolumn{17}{c}{KV Budget = 128 } \\
        StreamingLLM & 13.57 & 11.48 & 24.47 & \textbf{34.25} & \textbf{25.35} & \textbf{11.31} & \textbf{20.06} & \textbf{17.39} & 17.80 & \textbf{30.50} & 50.14 & 35.33 & \textbf{3.00} & 5.50 & 15.50 & 60.25 & 23.49\\
        Pyramid & \textbf{13.91} & \textbf{13.05} & \textbf{18.71} & 27.43 & 14.36 & 6.35 & 17.93 & 17.25 & 19.52 & 21.00 & 89.09 & \textbf{38.03} & 1.11 & 5.00 & 61.93 & 57.00 & 26.35 \\
        SnapKV & 12.23 & 11.88 & 17.93 & 25.32 & 12.41 & 8.20 & 17.88 & 16.84 & 19.44 & 22.00 & 90.97 & 37.83 & 2.50 & \textbf{5.50} & 61.62 & \textbf{57.54} & 26.26 \\
        Ada-SnapKV & 13.52 & 12.79 & 18.30 & 28.94 & 14.13 & 9.35 & 19.04 & 17.26 & \textbf{19.94} & 25.00 & \textbf{91.44} & 37.96 & 1.54 & 4.00 & \textbf{63.34} & 56.39 & \textbf{27.06} \\
        \midrule
        \multicolumn{17}{c}{KV Budget = 256 } \\
        StreamingLLM & 14.25 & 12.92 & \textbf{28.75} & 31.80 & \textbf{19.19} & \textbf{11.39} & \textbf{22.86} & 17.26 & 22.73 & \textbf{44.00} & 52.80 & 38.77 & 4.00 & 9.00 & 16.11 & 59.79 & 25.35 \\
        Pyramid & 13.57 & \textbf{17.36} & 19.29 & 30.95 & 18.10 & 8.21 & 20.30 & 17.65 & 20.90 & 26.00 & 90.94 & 40.02 & 4.73 & 4.50 & 64.71 & 55.06 & 28.27 \\
        SnapKV & 14.82 & 16.32 & 18.44 & 29.53 & 14.25 & 9.26 & 20.32 & 17.56 & 21.70 & 28.50 & 91.49 & 40.15 & 4.12 & 5.00 & 64.60 & \textbf{54.78} & 28.18 \\
        Ada-SnapKV & \textbf{15.19 }& 15.97 & 20.05 & \textbf{35.08} & 16.31 & 8.32 & 21.31 & \textbf{17.99} & \textbf{22.11} & 33.00 & \textbf{91.68} &\textbf{ 41.28} & \textbf{5.93} & \textbf{5.50} & \textbf{66.14} & 54.76 & \textbf{29.41} \\
        \midrule
        \multicolumn{17}{c}{KV Budget = 512 } \\
        StreamingLLM & 14.47 & 16.86 & \textbf{34.18} & 31.88 & 18.76 & \textbf{11.80} & \textbf{26.01} & 18.20 & \textbf{25.10} & \textbf{53.00} & 60.04 & 41.23 & 4.00 & 9.50 & 19.74 & \textbf{57.87} & 27.66 \\
        Pyramid & 15.67 & 21.59 & 23.56 & 35.63 & \textbf{24.36} & 9.77 & 22.18 & 19.05 & 23.25 & 34.00 & 91.11 & 42.44 & 4.61 & \textbf{17.50} & 65.87 & 54.41 & 31.56 \\
        SnapKV & 15.22 & 22.90 & 23.41 & 33.00 & 22.54 & 8.91 & 22.63 & 18.31 & 23.50 & 35.00 & 91.39 & 41.85 & 4.50 & 15.50 & \textbf{66.79} & 53.75 & 31.20 \\
        Ada-SnapKV & \textbf{17.72} & \textbf{24.22} & 25.38 & \textbf{38.00} & 23.47 & 9.25 & 23.55 & \textbf{18.91} & 23.82 & 43.00 & \textbf{92.14} & \textbf{42.81} & \textbf{6.34} & 15.00 & 66.12 & 55.32 & \textbf{32.82} \\
        \midrule
        \multicolumn{17}{c}{KV Budget = 1024 } \\
        StreamingLLM & 13.36 & 21.55 & \textbf{41.70} & 32.80 & 22.63 & 12.88 & 28.40 & 19.36 & \textbf{25.93} & \textbf{62.00} & 75.54 & 41.76 & 2.00 & 11.00 & 22.91 & \textbf{57.05} & 30.68 \\
        Pyramid & 16.67 & 29.11 & 28.74 & 40.13 & 27.89 & 13.43 & 24.63 & 20.36 & 25.01 & 43.00 & 91.86 & 43.48 & \textbf{6.78} & 52.50 & 65.34 & 54.87 & 36.49 \\
        SnapKV & 19.57 & 31.67 & 31.52 & \textbf{42.04} & 27.89 & 12.94 & 25.31 & 19.66 & 25.12 & 48.50 & 91.37 & 42.83 & 5.99 & \textbf{56.00} & \textbf{66.56} & 53.97 & 37.56 \\
        Ada-SnapKV & \textbf{19.74} & \textbf{30.47} & 31.42 & 40.82 & \textbf{28.98} & \textbf{16.07} & \textbf{25.35} & \textbf{20.57} & 25.57 & 53.50 & \textbf{91.76} & \textbf{43.81} & 4.63 & 53.50 & 65.75 & 55.90 & \textbf{37.99} \\
        \midrule
        \multicolumn{17}{c}{KV Budget = 2048 } \\
        StreamingLLM & 16.86 & 31.83 & \textbf{48.22} & 33.59 & 29.36 & 15.74 & \textbf{30.11} & 20.89 & 26.68 & \textbf{65.50} & \textbf{92.49} & 44.12 & 5.25 & 21.00 & 39.75 & \textbf{56.66} & 36.13 \\
        Pyramid & 21.16 & 36.66 & 37.56 & 43.28 & 36.25 & \textbf{19.72} & 27.90 & 21.73 & 26.35 & 53.00 & 92.36 & \textbf{44.61} & 6.94 & \textbf{89.50} & 64.23 & 53.11 & 42.15 \\
        SnapKV & 21.24 & 39.86 & 38.21 & 45.96 & 35.36 & 21.20 & 28.68 & 21.49 & 26.50 & 58.00 & 92.36 & 44.06 & 6.31 & 88.50 & \textbf{64.26} & 53.73 & 42.86 \\
        Ada-SnapKV & \textbf{22.42} & \textbf{40.26} & 40.13 & \textbf{49.60} & \textbf{38.23} & 19.70 & 28.49 & \textbf{21.76} & \textbf{26.77} & 61.50 & 92.35 & 44.04 & \textbf{8.27 }& 85.00 & 64.11 & 54.09 & \textbf{43.55} \\
        \bottomrule
    \end{tabular}
    \end{adjustbox}
    \caption{    Comparison of Common KV Cache Compression Methods.
    % All these methods were tested on the Llama-3.1-8B-Instruct using the LongBench v1 dataset, and the KV Cache budget is set to 128, 256, 512, 1024, and 2048 respectively.
    }
    \label{tab:KV_Methods}
\end{table*}

\section{Common KV Cache Compression Methods with Tensor Parallelism}

\begin{figure*}[htbp]
    \centering
    \includegraphics[width=\textwidth]{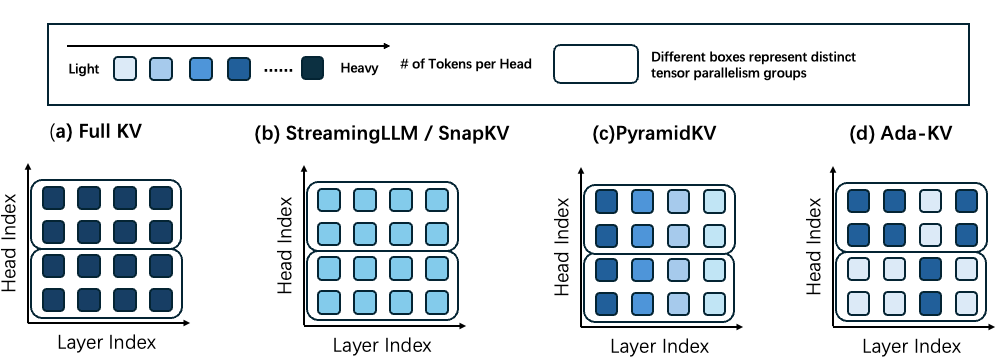}
    \caption{Common KV Cache Compression Methods with Tensor Parallelism}
    \label{fig:KV_methods_with_TP}
\end{figure*}

Figure~\ref{fig:KV_methods_with_TP} shows the combination of common KV cache compression methods with tensor parallelism. As can be seen from the figure, the traditional Balanced (Fair) Per-Head Compression methods result in a balanced computational load after the partitioning in tensor parallelism. In these methods, each part of the parallel computation bears a relatively equal amount of work. On the contrary, the Imbalanced (Unfair) Per-Head Compression methods lead to an unbalanced computational load.

\end{document}